\documentclass[12pt]{iopart}
\usepackage{graphicx}
\begin{document}
\def \lt{\!\!<\!}

\title[Tight cosmological constraints from the
angular-size/redshift relation]{Tight cosmological constraints
from the angular-size/redshift relation for ultra-compact radio
sources}

\author{J C Jackson\dag
\footnote[2]{To whom correspondence should be addressed
(john.jackson@unn.ac.uk)}}

\address{\dag\ Division of Mathematics and Statistics, School of Informatics,
Northumbria University, Ellison Place, Newcastle NE1 8ST, UK}

\begin{abstract}

Some years ago (Jackson and Dodgson 1997) analysis of the
angular-size/redshift relationship for ultra-compact radio sources
indicted that for spatially flat universes the best choice of
cosmological parameters was $\Omega_m=0.2$ and $\Omega_\Lambda=0.8$.
Here I present an astrophysical model of these sources, based upon
the idea that for those with redshift $z>0.5$ each measured angular
size corresponds to a single compact component which is moving
relativistically towards the observer; this model gives a reasonable
account of their behaviour as standard measuring rods.  A new analysis of the
original data set (Gurvits 1994), taking into account possible
selection effects which bias against large objects, gives
$\Omega_m=0.24+0.09/-0.07$ for flat universes. The data points 
match the corresponding theoretical curve very accurately
out to $z\sim 3$, and there is clear and sustained indication of
the switch from acceleration to deceleration, which occurs at
$z=0.85$.

\end{abstract}

\section{Introduction}

The default cosmological paradigm now is that we are living in a
spatially flat accelerating Universe with matter (baryons plus
Cold Dark Matter) and vacuum density-parameters $\Omega_m=0.27$
and $\Omega_\Lambda=0.73$ respectively, known as the concordance
model.  Definitive confirmation of a consensus which has been
growing over the last two decades came with the recent Wilkinson
Microwave Anisotropy Probe (WMAP) results (Spergel \etal 2003).
The original evidence for such models was circumstantial, in that
they reconcile the inflationary prediction of flatness with the
observed low density of matter (Peebles 1984; Turner \etal 1984).
The first real evidence came from observations of very large-scale
cosmological structures (Efstathiou \etal 1990), which paper
clearly advocated everything that has come to be accepted in
recent times: ``......very large scale cosmological structures can
be accommodated in a spatially flat cosmology in which as much as
80 percent of the critical density is provided by a positive
cosmological constant. In such a universe expansion was dominated
by CDM until a recent epoch, but is now governed by the
cosmological constant."  A similar case for this model was made by
Ostriker and Steinhardt (1995), who also noted that the location
and  magnitude of the first D\H oppler peak in the cosmic
microwave background (CMB) angular spectrum was marginally
supportive of flatness. However, the paradigm did not really begin
to shift until the Hubble diagram for Type Ia supernovae (SNe Ia)
(Schmidt \etal 1998; Riess \etal 1998; Perlmutter \etal 1999)
provided reasonably convincing evidence that $\Omega_\Lambda>0$;
the corresponding confidence region in the
$\Omega_m$--$\Omega_\Lambda$ plane was large and elongated, but
almost entirely confined to the positive quadrant.  The dramatic
impact of these results was probably occasioned by the simple
nature of this classical cosmological test; coupled with accurate
measures of the first D\H oppler peak in the CMB angular spectrum
(Balbi \etal 2000; de Bernardis \etal 2000; Hanany \etal 2000),
which established flatness to a high degree of accuracy, the SNe
Ia results made anything but the concordance model virtually
untenable.  This paper is in part retrospective, and is about
another simple classical cosmological test.  Some years ago we
published an analysis of the angular-size/redshift diagram for
milliarcsecond radio-sources (Jackson and Dodson 1997; see also
Jackson and Dodgson 1996), with a clear statement to the effect
that ``if the Universe is spatially flat, then models with low
density are favoured; the best such model is $\Omega_m=0.2$ and
$\Omega_\Lambda=0.8$''. This result pre-dates the SNe Ia ones.

Ultra-compact radio sources were first used in this context by
Kellermann (1993), who presented angular sizes for 79 objects,
obtained using very-long-baseline interferometry (VLBI).  These
were divided into 7 bins according to redshift $z$, and the mean
angular size $\theta$ plotted against the mean redshift for each
bin.  The main effect of Kellermann's work was to establish that
the resulting $\theta$--$z$ relationship was compatible with
standard Friedmann-Lema\^\i tre-Robertson-Walker (FLRW)
cosmological models, in sharp contrast to the case for the
extended radio structures associated with radio-galaxies and
quasars. In the latter case typical component separations are 30
arcseconds, and the observed relationship is the so-called
Euclidean curve $\theta\propto 1/z$ (Legg 1970; Miley 1971;
Kellermann 1972; Wardle and Miley 1974); this deficit of large
objects at high redshifts is believed to be an evolutionary
effect, brought about by interaction with an evolving
extra-galactic medium (Miley 1971; Barthel and Miley 1988; Singal
1988), or a selection effect, due to an inverse correlation
between linear size and radio power (Jackson 1973; Richter 1973;
Masson 1980; Nilsson \etal 1993).  However, see Buchalter \etal (1998)
for a significant attempt to disentangle these effects. 
Ultra-compact objects have short lifetimes and are much smaller than
their parent active galactic nuclei (AGNs), so that their local
environment should be free of cosmological evolutionary effects,
at least over an appropriate redshift range.  However, it is not
clear that observations of these objects are completely free from
selection effects.

Kellermann's work was extended by Gurvits (1994), who presented a
large VLBI compilation, based upon a 2.3 GHz survey undertaken by
Preston \etal (1985), comprising 917 sources with a correlated
flux limit of approximately 0.1 Jy; the sub-sample selected by
Gurvits comprises 337 sources with known redshifts, and objective
measures of angular size based upon fringe visibility (Thompson,
Moran and Swenson 1986).  Gurvits gave good reasons for
ignoring sources with $z<0.5$, and using just the high-redshift
data found marginal support for a low-density FLRW model, but
considered only models with $\Omega_\Lambda=0$. Jackson and
Dodgson (1997) was based upon Gurvits' sample, and considered 256
sources in the redshift range 0.511 to 3.787, divided into 16 bins
of 16 objects.  More recently Gurvits \etal (1999) have presented
a new compilation of 330 compact radio sources, observed at a
somewhat higher frequency ($\nu=5$ GHz), which has stimulated a
number of analyses (Vishwakarma 2001; Cunha \etal 2002; Lima 
and Alcaniz 2002; Zhu and Fujimoto 2002; Chen and Ratra 2003;
Jain \etal 2003), which consider the full $\Omega_m$--$\Omega_\Lambda$
plane and/or alternative models of the vacuum. However, the constraints
placed upon cosmological parameters by this later work are
significantly weaker than for example the SNe Ia constraints
alone, and very much weaker than those obtained when the latter
are coupled with CMB and Large Scale Structure observations
(Efstathiou \etal 1999; Bridle \etal 1999; Lasenby, Bridle and
Hobson 2000; Efstathiou \etal 2002).

The main purpose of this work is to show that
angular-size/redshift data relating to these sources place
significant constraints upon cosmological parameters, comparable
with those due to any of the currently fashionable tests taken in
isolation, and also to establish a plausible astrophysical model which
gives them credibility as putative standard measuring rods. I
find that the original Gurvits (1994) compilation better in this
respect than the later one due to Gurvits \etal (1999), and I
shall eventually discuss why this might be so.  The astrophysical model
and associated selection effects are discussed in Section 2, where
evidence of such an effect is found.  Appropriate countermeasures
are discussed in Section 3; these have some similarities to the scheme
adopted by Buchalter \etal (1998) with regard to the extended radio sources,
and the parallels will be discussed.  The corresponding cosmological
results are presented as marginalized confidence regions in the
$\Omega_m$--$\Omega_\Lambda$ plane, with some consideration of the quintessence
parameter $w$.  It is in these considerations that this work is new,
and differs significantly from the analysis of essentially the same data in
Jackson and Dodgson (1997).  At this stage I make no attempt to combine
these data with other observations, being content to show that the former
deserve to be part of the cosmological canon. Quoted figures which depend
upon Hubble's constant correspond to $H_0=100$ km sec$^{-1}$ Mpc$^{-1}$.

\section{Selection effects and a source model}

In a flux-limited sample sources observed at large redshifts are
intrinsically the most powerful, so that an inverse correlation
between linear size and radio luminosity will introduce a bias
towards smaller objects.  There are several reasons for expecting
such a correlation.  On quite general grounds we would expect
sources to be expanding, and their luminosities to decrease with
time after an initial rapid increase (Jackson 1973; Baldwin 1982;
Blundell and Rawlings 1999).  Additionally, D\H oppler beaming
from synchrotron components undergoing bulk relativistic motion
towards the observer is known to be important in compact sources,
and Dabrowski \etal (1995) have argued that this will introduce
a similar correlation.  D\H oppler boosting was first discussed by
Shklovsky (1964a,b), to explain the apparently one-sided jet in M87.
Subsequently Rees (1966) devised a relativistically expanding model
with spherical symmetry, to account for the rapid variability observed
in powerful radio sources, and noted that apparently superluminal
motion should be seen in such models, some ten years before this phenomen
was observed (Cohen 1975; Cohen \etal 1976, 1977). Shklovsky's basic
twin-jet model has been the subject many elaborations over the past
three decades, and D\H oppler boosting is now the basis of the so called
unified model of compact radio sources, in which orientation effects in
an essentially homogeneous population generate the full range of 
observed properties.  The notion that quasi-stellar radio sources
are just a small subset of apparently radio-quiet quasi-stellar objects,
that is those which are viewed in the appropriate orientation, was 
first discussed by Scheuer and Readhead (1979), and subsequently by
Orr and Browne (1982), who considered the relative frequencies of 
the two classes.  Particulary germane to the discussion below are
Blandford and K\H onigl (1979a), Blandford and K\H onigl (1979b) and
Lind and Blandford (1985).  This is not intended to be a comprehensive
historical review, examples of which can be found in Blandford \etal
(1977) and Kellerman (1994).

The underlying source population probably consists of compact
symmetric objects (CSOs), of the sort observed by Wilkinson \etal (1994)
at moderate redshifts ($0.2~^<_\sim~z~^<_\sim~0.5$), with radio
luminosity densities of several times $10^{26}$ W Hz$^{-1}$ at 5 GHz;
these comprise central low-luminosity cores straddled by two mini-lobes,
the former contributing no more than a few percent of the total luminosity
(Readhead \etal 1996a); without D\H oppler boosting these objects would be
too faint to be observed at higher redshifts.  It is thus reasonable to
suppose that in the most distantly observed cases the lobes are moving
relativistically and are close to the line of sight.  For this
reason the use of ultra-compact sources as standard measuring rods
has been questioned by Dabrowski \etal (1995), who consider a simple model,
comprising two identical but oppositely directed jets (treated as point or
line sources), and assume that the measured angular size corresponds to their
separation projected onto the plane of the sky; thus apparent radio power
increases and angular size decreases as the beams get closer to the line of sight.
However, this model is not realistic, as the counter-jet would be very much
fainter than the forward one, for example by a factor of up to $10^6$
for a jet Lorentz factor $\gamma$ of $5$. It is more reasonable to
suppose that we observe just that component which is moving
relativistically towards the observer, and in particular that the
interferometric angular sizes upon which this work is based
correspond to the said components.  

It is reasonable to ask if the above supposition is compatible with VLBI
images of distant AGNs and quasars.  At first sight this is not the case;
it is well-known that these images typically show a core/one-sided jet
structure (see for example Taylor \etal 1996), rather than a single component.
However, it is well-known that there is an inconsistency here, if the cores 
are to be identified with those of the underlying CSO population and the 
latter are unbeamed.  The problem then is that as mentioned above members 
of the CSO population are typically jet dominated, and would become distinctly 
more so when viewed close to the jet axis, by several orders of magnitude,
which is not what is observed; in VLBI images of distant sources the core 
is usually dominant, and the jet is often absent.  This dilema has been 
resolved by Blandford and K\H onigl (1979a,b), who describe a model in which 
the `core' is really part of the jet, and is thus also D\H oppler boosted
(see also Kellerman 1994).  In their model the core emmission originates in
the stationary compact end of a quasi-steady supersonic jet, where the latter
becomes optically thick; the so called jet emmission comes from up-stream shock
waves associated with dense condensations within the bulk flow which are being
accelerated by the latter.  A stationary core which is nevertheless relativistic
is necessary, to reconcile superluminal expansion with the observed relative fluxes
from the two components.  In the latter respect it is essential that $\gamma_{\hbox{\small{j}}}<\gamma_{\hbox{\small{c}}}$
(where subscripts c and j denote core and jet respectively), but this is part of
the Blandford and K\H onigl (1979a,b) model; jet domination then changes into core
domination as $\phi$ changes from $\pi/2$ to zero, where $\phi$ be the angle between
the jet axis and the line-of-sight; the transverse D\H oppler effect diminishes the
core relative to the jet when $\phi=\pi/2$, but the roles are reversed when $\phi=0$.
In other words the core emission is more beamed than that of the jet.

If the two components have flat spectra and the same rest-frame luminosities,
and $R(\phi)$ is the core/jet luminosity ratio, then

\begin{equation}\label{R}
R(0)
=\left({\gamma_{\hbox{\small{c}}} \over \gamma_{\hbox{\small{j}}}}\right)^n
~~~~~~\hbox{and}~~~~~~
R(\pi/2)
=\left({\gamma_{\hbox{\small{j}}} \over \gamma_{\hbox{\small{c}}}}\right)^n
\end{equation}

\noindent
and the crossover angle at which $R=1$ is
$\phi\sim (\gamma_{\hbox{\small{j}}}\gamma_{\hbox{\small{c}}})^{-1/2}$
(see below for the definition of $n$).  A ratio $\gamma_{\hbox{\small{c}}}/\gamma_{\hbox{\small{j}}}\sim 2\hbox{ to }3$
would effect a transition of the correct magnitude; for typical values
of $\gamma$ the crossover angle is $15^\circ$ to $20^\circ$. 

If the basic model outlined above is correct, one consequence is that
statistically cores are observed at something like a fixed rest-frame frequency;
suppose that a core has rest-frame luminosity density $L$, with flat spectral
index $\alpha=0$ characteristic of ultra-compact sources.  The D\H oppler
boosting factor ${\cal D}$ is

\begin{equation}\label{D}
{\cal D}=\gamma^{-1}(1-\beta\cos\phi)^{-1}    
\end{equation}

\noindent
where $\beta$ is the object velocity in units of $c$, and as above
$\phi$ is the angle between this velocity and the line of sight.
For a source at redshift $z$ and angular-diameter distance $D_A(z)$,
the observed flux density $S$ is thus

\begin{equation}\label{S}
S={L{\cal D}^n \over 4\pi(1+z)^nD_A^2}
\hbox{\quad}\Rightarrow\hbox{\quad}{{\cal D}\over 1+z}=\left({4\pi D_A^2S
\over L}\right)^{1/n}(1+z)^{(3-n)/n}
\end{equation}

\noindent
where $n=3$ for discrete ejecta and $n=2$ for a continuous jet
(Lind and Blandford 1985).  As we have seen, in reality the situation
lies between these two extremes, and a value $n=5/2$ will be used for
purposes of illustration.  In the redshift range of interest $D_A(z)$
is close to its minimum, and is thus a slowly varying function of $z$,
so that for an object observed close to the flux-limit the ratio
${\cal D}/(1+z)$ is proportional to $(1+z)^{1/5}$ and is roughly fixed:
the survey frequency of 2.3 GHz corresponds to a rest-frame frequency
of 2.3 GHz divided by this ratio, which would for example greatly reduce
the effect, deleterious in this context, of any dependence of linear size on
rest-frame frequency.  Similar considerations apply to jets if
$\gamma_{\hbox{\small{j}}}/\gamma_{\hbox{\small{c}}}$ is fixed.

It has been noted by Frey and Gurvits (1997) that jets become noticeably less
prominent, in terms of both morphology and luminosity, as redshift increases.
At $z>3$ jets appear to be absent or vestigial (see for example the VLBI images
presented by Gurvits \etal 1994, Frey \etal 1997 and Paragi \etal 1999).
Frey and Gurvits (1997) reasonably attribute this phenomenon to differential
spectral properties, cores being flatter than the jets in this respect.
However, if statistically the cosmological redshift is roughly cancelled out
by the D\H oppler boost, as suggested in the last paragraph, then spectral
differences should be not be as important as Frey and Gurvits (1997) suppose.
In part the phenomenon is probably the Dabrowski \etal (1995) selection effect
in operation: the viewing angle $\phi$ gets smaller as $z$ increases, and projecion
effects mean that the two components first overlap and then become superimposed,
when we see a single composite source.  Such selection is thus not as significant
as Dabrowski \etal (1995) suggest, particularly at high redshifts, because the
components are not point sources, and angular sizes do not vanish as the beams
get closer to the line of sight. At lower redshifts we see more structure;
cursory inspection of 113 VLBI images (not redshift selected) presented by
Taylor \etal (1996) suggests that about one third of these are superimposed
composites, one third are core/jet overlaps, and one third show two separate
components or more complex structures.  However, Dabrowski \etal (1995) show
that their effect is not significant at low redshifts, typically
$z~^<_\sim~1.5$ for the flux limit of 0.1 Jy which characterises this sample.
Thus over the full redshift range it is reasonable to suppose that this particular
selection is of marginal importance.  The matter will be discussed further,
when suitable countermeasures are considered.  Although these considerations
are interesting from an astrophysical point of view, their significance here
is to identify these superimposed core/jet composites as the components which
in effect determine the measured angular sizes upon which this work is based.  
     
Finally, with regard to establishing ultra-compact sources as standard
linear measures, I note that the parent galaxies in which they are
embedded are giant ellipticals, with masses close $10^{13}M_\odot$;
it is known that their central black holes have masses which are tightly
correlated with this mass (Kormendy 2001a,b), being 0.15\% of
same, close to $1.5\times 10^{10}M_\odot$.  The central engines
which power these sources are thus reasonably standard objects.

\begin{figure}[here]
\begin{center}
\includegraphics[width=12.5cm] {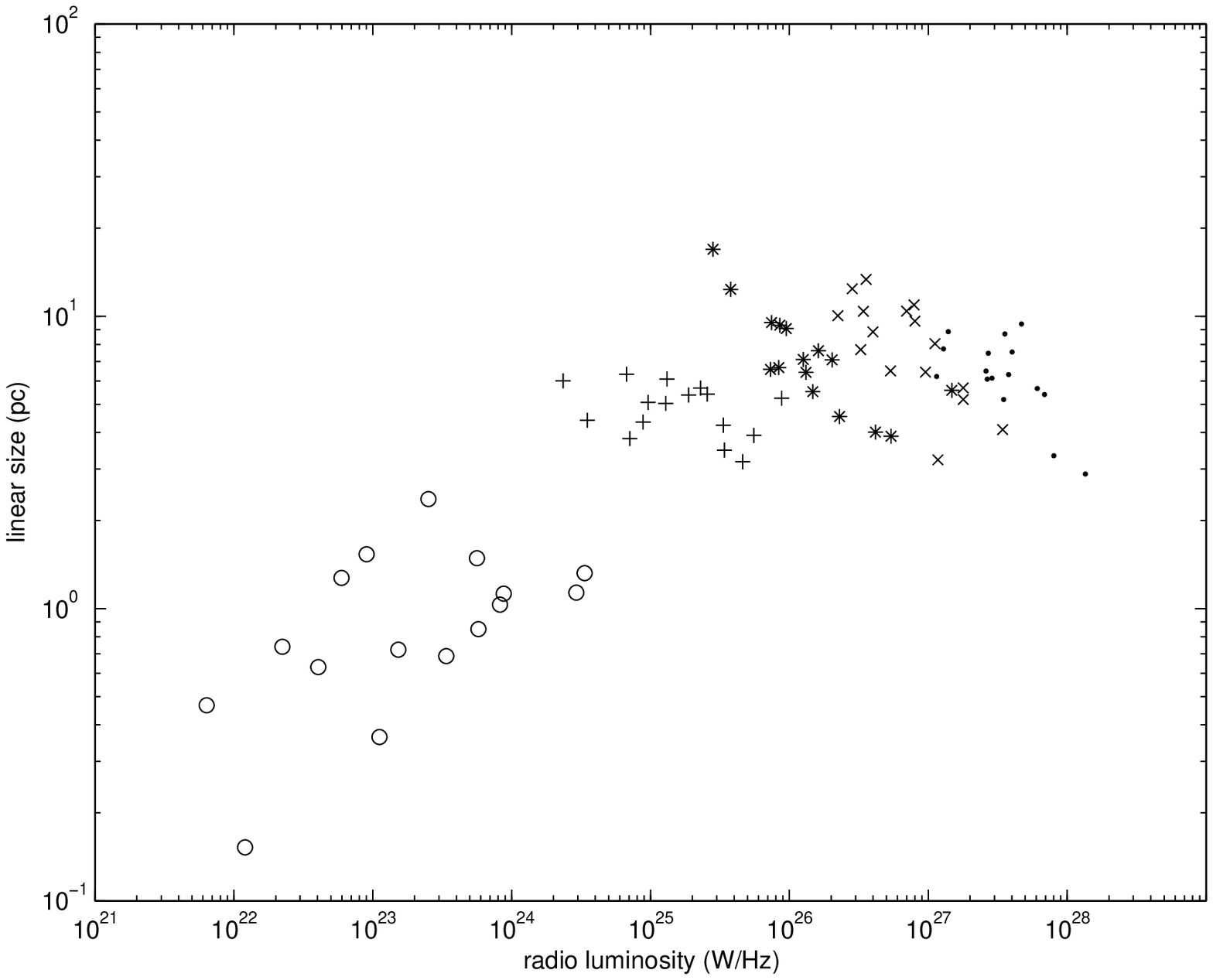}
\end{center}
\caption{\label{Fig1}Linear size versus radio luminosity for sources in selected redshift bins: 
$\circ\; 0.00\lt z\lt 0.06$, $+\; 0.21\lt z\lt 0.31$, $\ast\; 0.51\lt z\lt 0.58$,
$\times\; 1.15\lt z\lt 1.29$, $\cdot\; 2.70\lt z\lt 3.79$, showing the inverse correlation between linear size and radio power at high redshifts.}
\end{figure}

The question of whether there is a linear-size/luminosity correlation
of the sort discussed  at the beginning of this section is easily settled;
Figure \ref{Fig1} is a plot of linear extent $d$ (as indicated by the measured
angular size) against correlated rest-frame luminosity $L$ (attributable to
the compact component, and calculated assuming isotropic emission and a spectral
index $\alpha=-0.1$, where $L\propto\nu^{-\alpha}$), for a selection of 
redshift bins, each containing 16 sources from the Gurvits (1994) sample.
A cosmology is needed for this plot, and I have pre-empted the results to be
presented here by using $\Omega_m=0.24$ and $\Omega_\Lambda=0.76$; however,
the significant qualitative aspects of the arguments I shall present are not
sensitive to this choice.  There is clear evidence that at any particular
epoch individual luminosities are a function of size, and that for sources
with $z~^>_\sim~0.2$ this relationship is an inverse correlation.
For $z>0.5$ the relationship is well-represented by

\begin{equation}\label{L}
L~\propto~d^{-a}
\end{equation}

\noindent
where in round figures $a=3$.  In the above model, this behaviour
would be attributed to expansion of each relativistic component as
it moves away from the central engine and grows weaker.  However,
as noted by Gurvits (1994), the remarkable feature of this diagram
is that for $z>0.5$ there is no marked change in mean size at a
given luminosity, over a range of the latter variable which covers
three orders of magnitude, which suggests that these two
parameters are decoupled.  This behaviour is quite compatible with
the model outlined above; a tentative subdivision is that for
$z<0.2$ these sources are intrinsically small and weak and not
relativistic; between $z=0.2$ and $z=0.5$ there is a transitional
regime, during which the intrinsic luminosity rises to several
times $10^{26}$ W Hz$^{-1}$; thereafter the latter is approximately
fixed, the sources are ultra-relativistic, and the luminosity
range is accounted for largely by changes in D\H oppler boost.
For example with fixed $\gamma=5$, a luminosity boost factor
${\cal D}^{2.5}$ varies by a factor of $316$ as $\phi$ changes from
$0^\circ$ to $35^\circ$.  A similar phenomenon has been noted in
X-ray astronomy (Fabian \etal 1997).

The sub-division outlined above is analogous to the division of extended
double radio sources into Fanaroff-Riley types I and II (FR-I and FR-II)
(Fanaroff and Riley 1974; Kembhavi and Narlikar 1999).  The radio morphologies
of FR-I sources typically show a relatively bright central core from which
bright opposed jets emanate in symmetric fashion, which jets terminate in faint
diffuse lobes as they run into the inter-galactic medium.  FR-II sources typically
comprise a relatively faint core with a one-sided jet which terminates in a bright
compact lobe, with a comparable lobe on the opposite side of the core, but no
sign of a corresponding counter-jet; presumably the latter is invisible due to
beaming.  FR-I sources are relatively weak, having radio luminosities
$L(\hbox{2.3 GHz})~^<_\sim~2 \times 10^{24}$ W Hz$^{-1}$, whereas FR-II sources
are intrinsically strong, having luminosities greater than this figure.
Although this threshold luminosity is two orders of magnitude lower than the
luminosities associated with the CSOs observed by Wilkinson \etal (1994), it may 
be that the analogy is exact, in that latter are the precursors of FR-I objects,
if luminosity evolution is invoked (Readhead \etal 1996b).  Similarly, the more 
distant compact objects discussed here may be the precursors of FR-II sources. 

Buchalter \etal (1998) select FR-II sources as the basis of their work on
the angular-size/redshift, relation, in part because their morphology allows
an objective definition angular size.  This selection was effected by choosing
sufficiently powerful sources having $z>0.3$ and the correct radio morpology.
Buchalter \etal (1998) must then first allow for a lower size cut-off, below
which the FR-I/FR-II classification is beyond the resolving power of the
instruments in question.  Finally Buchalter \etal (1998) allow for a
linear-size/redshift correlation of the sort discussed above, by assuming
$d~\propto~(1+z)^c$ and allowing the data to fix the constant $c$, which
represents a combination of several effects.  Such measures appear to be
necessary, if the angular-size/redshift diagram for extended sources is
to be in concordance with acceptible FLRW cosmological models, but the situation
is then too degenerate to make definitive statements about cosmological parameters.
As we shall see in the next section, milliarcseconds sources are more robust in this
respect, and although similar  measures will be introduced, they are a refinement
rather than a necessity.

Because we cannot measure absolute visual magnitudes or linear
sizes directly, it is essential to have a model which supports the
choice of the objects in question as standard candles or standard
measuring rods. In the case of Type Ia supernovae such support is
provided by a reasonably well-established model, based upon
accreting Chandrasekhar-mass white dwarfs in binary systems
(Branch \etal 1995; Livio 2001).  I believe that the above model
provides similar support for powerful ultra-compact radio sources
and the angular-size/redshift diagram.

\section{Cosmological parameters}

As in Gurvits (1994) and Jackson and Dogson (1997), I shall
consider only those sources with $z>0.5$, amply justified by the
discussion above.  It is clear that as individual objects these do
not have fixed linear dimensions, and we must consider the
population mean.  Accordingly, the sources are placed in redshift
bins, and earlier practice would be to plot simple means for
each bin in a $\theta$--$z$ diagram.  However, due to the
selection biases discussed in Section 2, a growing proportion of the
larger members of this population will be lost as redshift
increases, and simple means introduce a systematic error.
According to the adopted model, the effect can be quantified. The
flux limit $S_c$ sets a cut-off size $d_c$ at each redshift, such
that sources larger than $d_c$ are too faint to be observed. Again
taking a flat spectrum, equations (\ref{S}) and (\ref{L}) with $a=3$ give

\begin{equation}\label{d_c1}
d_c~\propto~{{\cal D}^{n/3} \over (4\pi D_A^2S_c)^{1/3}(1+z)}.
\end{equation}

\noindent
Thus assuming that there is a largest boost factor ${\cal D}$
which does not change with redshift, and noting that in the
redshift range of interest $D_A(z)$ is close to its minimum, we
find

\begin{equation}\label{d_c2}
d_c~\propto~(1+z)^{-1}.
\end{equation}

\begin{figure}[here]
\begin{center}
\includegraphics[width=12.5cm] {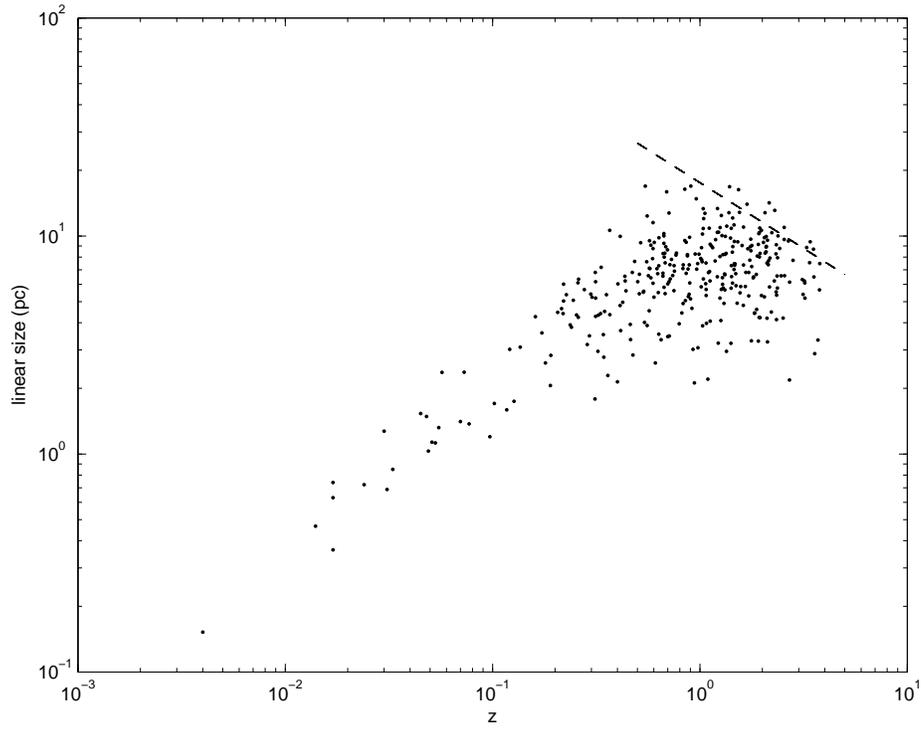}
\end{center}
\caption{\label{Fig2}Linear size versus redshift for the 337 sources in the sample used here.
The dashed cut-off curve corresponds to the model discussed in the text, in which larger sources are intrinsically weaker.}
\end{figure}

\begin{figure}[here]
\begin{center}
\includegraphics[width=12.5cm] {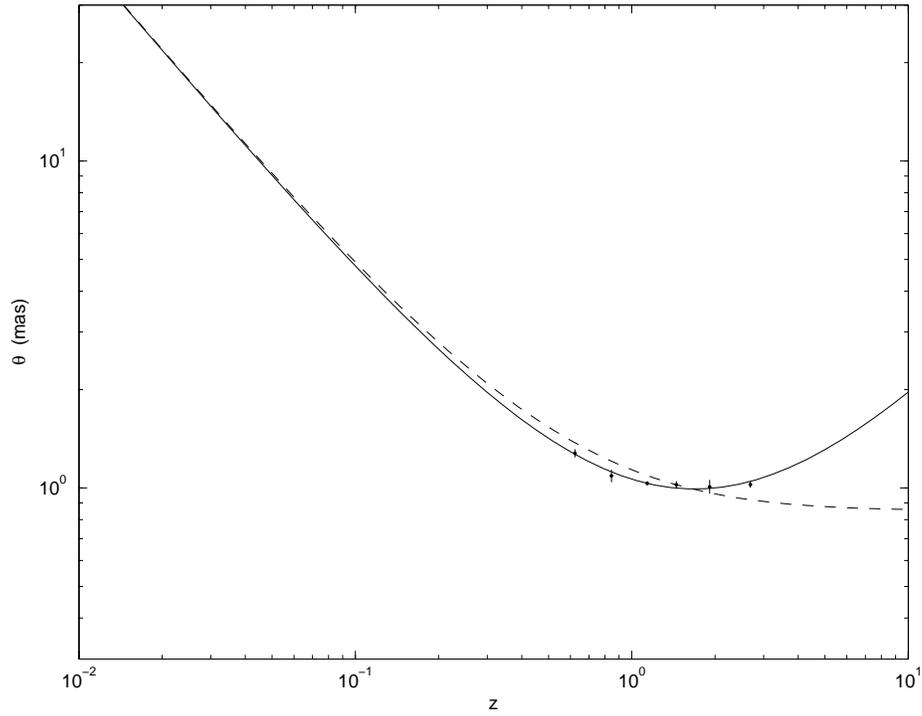}
\end{center}
\caption{\label{Fig3}Angular size $\theta$ versus redshift for 252 sources with
$0.541\leq z\leq 3.787$, divided into 6 bins of 42 objects; $\full\; \Omega_m=0.24$, $\Omega_\Lambda=0.76$; $\dashed\; \Omega_m=0$, $\Omega_\Lambda=0$.  The angular size
corresponds roughly to the 14th object within each bin; this definition is used to reduce the effects of bias against large objects, see text.}
\end{figure}

\noindent
Figure \ref{Fig2} is a plot of linear extent against $z$ for all 337
sources in the Gurvits (1994) sample. The dashed line shows
$d_c(z)$ according to equation (\ref{d_c2}), normalized to give $d_c=20$ pc
at $z=1$.  Figure \ref{Fig2} appears to show that the larger objects are
being lost when $z$ exceeds $1.5$, in a manner which is in reasonable 
accord with equation (\ref{d_c2}).  I have adopted the following pragmatic
procedure to reduce the concomitant systematic error; instead of simple
means I have defined a lower envelope for the data.  An obvious choice
would be the lowest point within each bin, but this is too noisy.
I have defined the lower envelope as the boundary between the bottom
third and the top two thirds of the points within each bin, in other
words a median which gives more weight to the smaller objects.
Note however that this prescription is not tied to the particular
model of bias discussed in Section 2; it is an empirical measure
based upon the observation that larger sources appear to be weaker,
and its effect would be neutral otherwise.  (Gurvtis \etal (1999)
use ordinary median angular sizes rather than means, to reduce 
the influence of outliers, which is an added benefit here.)
Figure \ref{Fig3} shows 6 weighted median points; each of these derives
from a bin containing 42 objects, being the mean of points 11 to 17
within each bin, counting from the smallest object; the sample thus
comprises 252 objects in the range $0.541\leq z\leq 3.787$.  Error
bars are $\pm$ one standard deviation as determined by the said points;
they are shown as an indication of the efficacy of this definition
of lower boundary, and are not used in the statistical analysis
which follows.  I have experimented with various bin sizes; the
reasons for this particular choice will be discussed later.  In
all cases means are means of the logarithms.

\begin{figure}[here]
\begin{center}
\includegraphics[width=12.5cm] {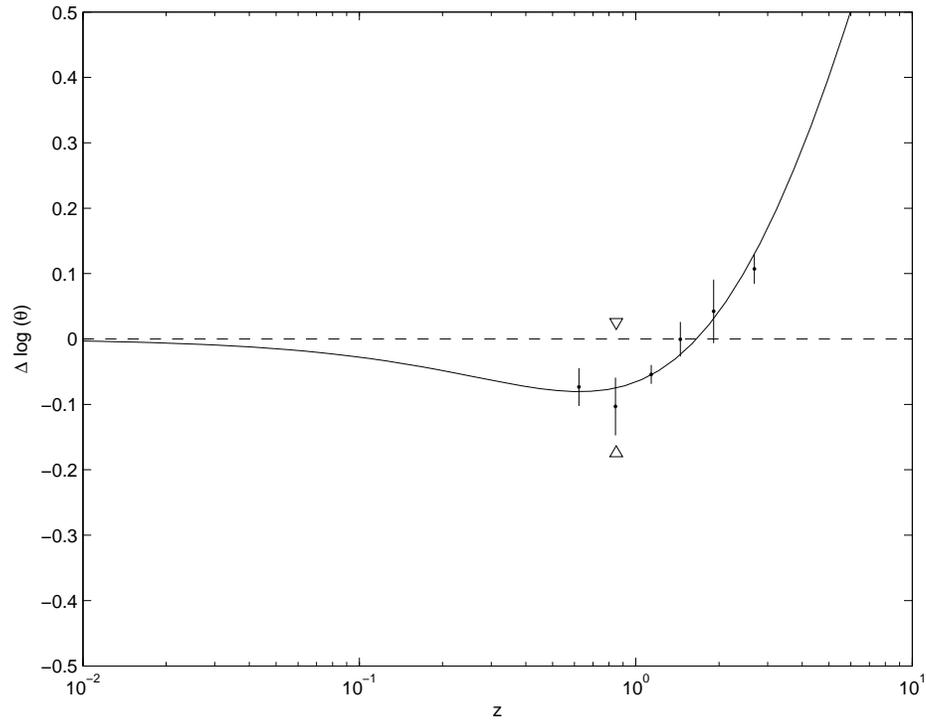}
\end{center}
\caption{\label{Fig4}Differential form of Figure 3; $\full\; \Omega_m=0.24$, $\Omega_\Lambda=0.76$; $\dashed\; \Omega_m=0$, $\Omega_\Lambda=0$; the triangles 
delineate the transition from acceleration to deceleration, at $z=0.85$.}
\end{figure}

A simple three-parameter least-squares fit to the points in Figure
\ref{Fig3} gives optimum values $\Omega_m=0.29$, $\Omega_\Lambda=0.37$ and
mean size $d=5.7$ parsecs. If the model is constrained to be flat,
then a two-parameter least-squares fit gives optimum values
$\Omega_m=0.24$, $\Omega_\Lambda=1-\Omega_m=0.76$ and $d=6.2$
parsecs, which model is shown as the continuous curve in Figure 3;
the latter also shows the zero-acceleration model $\Omega_m=0$,
$\Omega_\Lambda=0$, $d=6.2$ pc as the dashed curve.  The
difference $\Delta\log(\theta)$ between the two curves is presented in
Figure \ref{Fig4}, which shows clearly the shift from acceleration to
deceleration.  Note however that the actual switch occurs before
the crossing point, at $z=(2\Omega_\Lambda/\Omega_m)^{1/3}-1=0.85$
in this case, roughly where the continuous curve begins to swing
back towards the dashed one. Figure \ref{Fig4} establishes definitively and
accurately that there is no need to invoke anything other than a
simple $\Omega_m$--$\Omega_\Lambda$ model to account for the data,
out to a redshift $z=2.69$.  The current record for SNe Ia is
SN 1997ff at $z\sim 1.7$ (Gilliand and Phillips 1998; Riess \etal 2001),
with a somewhat uncertain apparent magnitude.

\begin{figure}[here]
\begin{center}
\includegraphics[width=12.5cm] {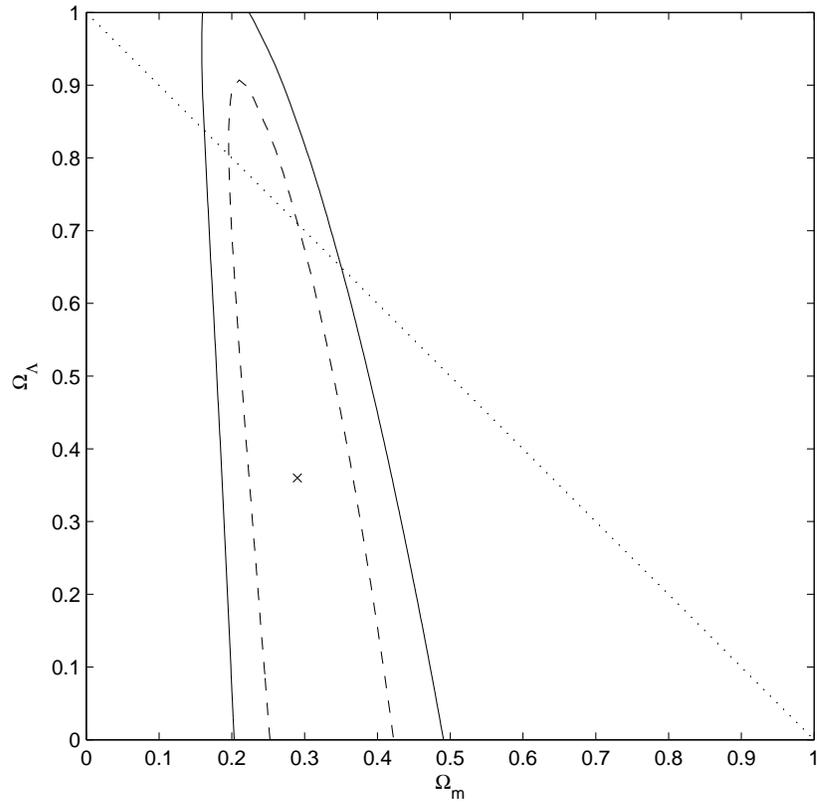}
\end{center}
\caption{\label{Fig5}Confidence regions in the $\Omega_m$--$\Omega_\Lambda$ plane, marginalised
over the linear dimension $d$; $\full\; 95\%$, $\dashed\; 68\%$, $\dotted\;$ flat.
The cross indicates the global minimum in $\chi^2$, at $\Omega_m=0.29$, $\Omega_\Lambda=0.37$.}
\end{figure}

\begin{figure}[here]
\begin{center}
\includegraphics[width=12.5cm] {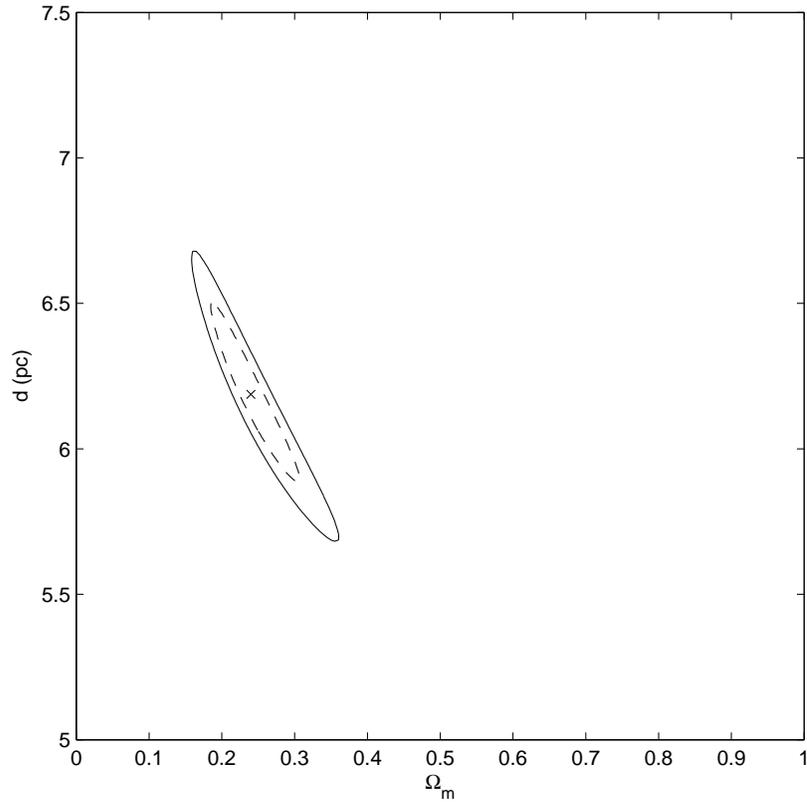}
\end{center}
\caption{\label{Fig6}Confidence regions in the $\Omega_m$--$d$ plane for flat models;
$\full\; 95\%$, $\dashed\; 68\%$.  The cross indicates the global minimum in $\chi^2$,
at $\Omega_m=0.24$, $d=6.20$ pc.  Marginalising over the linear dimension $d$ gives $\Omega_m=0.24+0.09/-0.07$.}
\end{figure}

In order to derive confidence regions, I have defined a fixed
standard deviation $\sigma$ to be attached to each point in Figure
\ref{Fig3}: $\sigma^2=\hbox{residual sum-of-squares/}(n-p)$, where
$n=6$ is the number of points and $p=3$ is the number of fitted
parameters, giving $\sigma=0.0099$ in log $\theta$ in this case.
This value of $\sigma$ is used to calculate $\chi^2$ values at
points in a suitable region of parameter space.  Figure 5 shows
confidence regions in the $\Omega_m$--$\Omega_\Lambda$ plane derived
in this manner, marginalized over $d$ according to the scheme
outlined in Press \etal (1986). Without the extra constraint of
flatness little can be said about $\Omega_\Lambda$, which degeneracy
is due to the lack of data points with $z<0.5$ (Jackson and 
Dodgson 1996). Nevertheless Figure \ref{Fig5} clearly constrains
$\Omega_m$ to be significantly less than unity. Figure \ref{Fig5} is
essentially a refined and tighter version of the diagram presented
in Jackson and Dodgson (1997); the significant change is that flat
models are now well within the 68\% confidence region, whereas
previously the figure was 95\%; this change is entirely due to the
measures relating to selection effects.  In the case of flat
models, two-dimensional confidence regions can be presented
without marginalization. Figure \ref{Fig6} shows such regions in the
$\Omega_m$--$d$ plane; marginalizing over $d$ gives 95\%
confidence limits $\Omega_m=0.24+0.09/-0.07$.

With respect to choice of redshift binning the balance is between
many bins containing few objects and a small number containing
many objects; the former gives poor estimates of population
parameters within each bin, but a large number of points to work
with in Figure \ref{Fig3}; the latter gives better estimates of these
parameters but fewer points in Figure \ref{Fig3}. I have experimented with
15 bin sizes, from 17 bins of 15 objects to 4 bins of 57 objects,
in steps of 3 objects; in each case I have used the largest number
of bins compatible with having no objects with $z\leq 0.5$. I find
that the central figure is quite robust, but that accuracy
increases gradually as the number of bins is reduced, the best
compromise being the one used above. As a check I have calculated
the best cosmological parameters in each of the 15 cases, and when
selection effects are allowed for I find a mean value
$\Omega_m=0.25$ in the flat case, with 95\% confidence limits of
$\pm 0.06$. If selection effects are ignored corresponding figures
are virtually the same, $\Omega_m=0.24\pm 0.04$, again in the flat
case.  However, this coincidence understates the value of the
measures relating to selection bias; as mentioned above, these
bring flat models well within the 68\% confidence region in Figure
5, and more importantly they reveal the expected minimum angular
size in Figure \ref{Fig3} (Hoyle 1959).

\begin{figure}[here]
\begin{center}
\includegraphics[width=12.5cm] {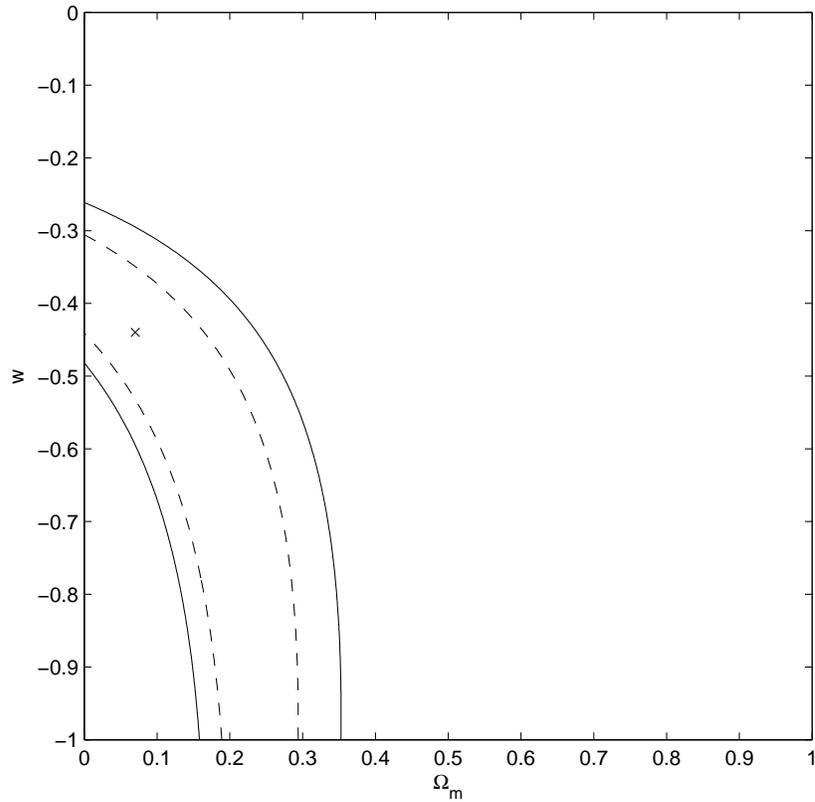}
\end{center}
\caption{\label{Fig7}Confidence regions in the $\Omega_m$--$w$ plane for flat models,
marginalised over the linear dimension $d$, where $w$ is the quintessence parameter;
$\full\; 95\%$, $\dashed\; 68\%$.  The cross indicates the global minimum in $\chi^2$,
at $\Omega_m=0.07$, $w=-0.44$.}
\end{figure}

As already noted with respect to Figure \ref{Fig4}, simple
$\Omega_m$--$\Omega_\Lambda$ models give an excellent fit to the
data.  Nevertheless, in conclusion I consider the limits placed
upon the quintessence parameter $w$, defined by postulating an
equation of state for the vacuum of the form
$\rho_{\hbox{\scriptsize vac}}=wp_{\hbox{\scriptsize vac}}$
relating the vacuum density $\rho_{\hbox{\scriptsize vac}}$ to the
vacuum pressure $p_{\hbox{\scriptsize vac}}$, with $|w|\leq 1$ and
$w=-1$ corresponding to the conventional vacuum defined by a
cosmological constant.  For flat models we have a three-parameter
system comprising $\Omega_m$, $w$ and $d$, the quintessence
parameter $\Omega_q$ being $1-\Omega_m$; we proceed by
marginalising over $d$ to give the two-parameter confidence
regions shown in Figure \ref{Fig7}.  The system is highly degenerate, and
with respect to material content cannot distinguish between
between for example a two component mix with $\Omega_m=0.24$,
$\Omega_q=0.76$, $w=-1$ at one extreme, and a single component
compromise with $\Omega_m=0$, $\Omega_q=1$, $w=-0.37$ at the
other.  Lacking any compelling evidence to the contrary, the
sensible choice is to retain local Lorentz invariance and assume
that $w=-1$.

\section{Conclusions}

The prescription adumbrated here has produced a set of data points
which are remarkably consistent with $\Omega_m$--$\Omega_\Lambda$
FLRW cosmological models, but there is extensive degeneracy due to
the restricted redshift range.  This degeneracy is resolved by
combining angular-size/redshift data with that CMBR information
which indicates flatness, and the two data sets together give
$\Omega_m=0.24+0.09/-0.07$.  This compares well with the figure
$\Omega_m=0.27\pm 0.04$ arising from WMAP measurements combined
with a host of other astronomical data sets (Spergel \etal 2003);
the points generated in this work might be added as an extra data set;
for future reference these are given in Table \ref{Table}.

\begin{table}[here]
\caption{\label{Table}Data points for the angular-size/redshift relationship;
$\theta$ is in milliarcseconds.}
\begin{indented}
\item[]\begin{tabular}{@{}cc}
\br
$z$ & $\theta$ \\
\mr
0.623 & 1.277 \\
0.845 & 1.089 \\
1.138 & 1.034 \\
1.450 & 1.023 \\
1.912 & 1.008 \\
2.686 & 1.024 \\
\br
\end{tabular}
\end{indented}
\end{table}

This work builds upon the earlier work of Jackson and Dodgson
(1997), and shows that the results obtained there were not
spurious.  Its purpose goes beyond that of showing compatibility
with more recent work, and suggests that building a much larger
angular-size/redshift data set for ultra-compact sources would be
a promising enterprise.  VLBI resolution of a quasar at $z=5.82$
has been demonstrated (Frey \etal 2003), so that the
redshift limit of such a data set should go well beyond that 
expected of the Supernova/Accleration probe (SNAP) (Aldering
\etal 2004), approaching 6 rather than 1.7.  Additionally, this
approach is immune to effects which might invalidate the SNe Ia
results, such as absorption by grey dust, as has been noted
by others (Bassett and Kunz 2004a; Bassett and Kunz 2004b).
The two approaches are of course complementary in their redshift
ranges, rather than competitive.  Section 2 might act as a guide
to further work, particularly with regard to the reduction of
selection effects; additionally, samples might be filtered using
morphological considerations, to include only those objects which
show roughly circular symmetry, corresponding to the superimposed
core/jet composites discussed in Section 2, which procedure would
be similar to that adopted by Buchalter \etal (1998) in their
selection of FR-II sources.  A related proposal due to Wiik and
Valtaoja (2001) is that the linear sizes of shocks within jets
might be estimated directly for each object, using flux density
variations and light travel time arguments, so that each object
would become a separate point in the angular-size/redshift diagram;
a weakness in their case is that individual D\H oppler boosts have
to be estimated.

I must end on a cautionary note; in general results from
the 5GHz sample due to Gurvits \etal (1999) are compatible with the
ones presented here, but with much greater uncertainty (Chen and Batra 2003);
the prescription developed here does not significantly improve matters in this
respect.  The probable reason for this difference in behaviour is
the definition of angular size; in Gurvits \etal (1999) this is
defined as the distance between the strongest component and the most
distant one with peak brightness $\geq$ 2\% of that of the strongest
component; here the objective measure based upon fringe visibility
ignores such outliers and estimates the size of the strongest component.

\References

\item[] Aldering G \etal 2004 {\it Preprint} astro-ph/0405232 
\item[] Balbi A, Ade P, Bock J, Borrill J, Boscaleri A, De Bernardis P, Ferreira P G,
        Hanany S, Hristov V, Jaffe A H, Lee A T, Oh S, Pascale E, Rabii B, Richards P L,
        Smoot G F, Stompor R, Winant C D and Wu J H P 2000 { \it Astrophys. J.}
        {\bf 545} L1
\item[] Baldwin J E 1982 {\it Proc. IAU Symp. 97} (Dordrecht: Reidel) p 21
\item[] Barthel P D and Miley G K 1988 {\it  Nature} {\bf 333} 319
\item[] Bassett B A and Kunz M 2004a {\it Astrophys. J.} {\bf 607} 661
\item[] Bassett B A and Kunz M 2004b {\it Phys. Rev.} D {\bf 69} 101305
\item[] Blandford R D, McKee C F and Rees M J, 1977 {\it  Nature} {\bf 267} 211
\item[] Blandford R D and K\H onigl A 1979a {\it Astrophys. Lett.} {\bf 20} 15
\item[] Blandford R D and K\H onigl A 1979b {\it Astrophys. J.} {\bf 232} 34
\item[] Blundell K M and Rawlings S 1999 {\it  Nature} {\bf 399} 330
\item[] Branch D, Livio M, Yungelson L R, Boffi F R, Baron E and Baron E 1995
        {\it Proc. Astronom. Soc. Pacific} {\bf 107} 1019
\item[] Bridle S L, Eke, V R, Lahav O, Lasenby A N, Hobson M P, Cole S, Frenk C S
        and Henry J P 1999 {\it Mon. Not. R. Astron. Soc.} {\bf 310} 565
\item[] Buchalter A, Helfand D J, Becker R H and White R L 1998 {\it Astrophys. J.}
        {\bf 494} 503
\item[] Chen G and Ratra B 2003 {\it Astrophys. J.} {\bf 582} 586
\item[] Cohen M H 1975 {\it Ann. N. Y. Acad. Sci.} {\bf 262} 428
\item[] Cohen M H, Moffet A T, Romney J D, Schilizzi R T, Seielstad G A, Kellermann K I,
        Purcell G H, Shaffer D B, Pauliny-Toth I I K and Preuss E 1976
        {\it Astrophys. J.} {\bf 206} L1
\item[] Cohen M H, Linfield R P, Moffet A T, Seielstad G A, Kellermann K I, Shaffer D B, 
        Pauliny-Toth I I K, Preuss E, Witzel A and Romney J D 1977 {\it Nature}
        {\bf 268} 405
\item[] Cunha J V, Alcaniz J S and  Lima J A 2002 {\it Phys. Rev.} D {\bf 66} 023520
\item[] Dabrowski Y, Lasenby A and Saunders R 1995 {\it Mon. Not. R. Astron. Soc.}
        {\bf 277} 753
\item[] de Bernardis P \etal  2000 {\it  Nature} {\bf 404} 955
\item[] Efstathiou G, Sutherland W J and Maddox S J 1990 {\it  Nature} {\bf 348} 705
\item[] Efstathiou G, Bridle S L, Lasenby A N, Hobson M P and Ellis R S 1999
        {\it Mon. Not. R. Astron. Soc.} {\bf 303} L47
\item[] Efstathiou G \etal 2002 {\it Mon. Not. R. Astron. Soc.} {\bf 330} L29
\item[] Fabian A C, Brandt W N, McMahon R G and Hook I M 1997 {\it Mon. Not. R. Astron. Soc.}
        {\bf 291} L5
\item[] Fanaroff B L and Riley J M 1974 {\it Mon. Not. R. Astron. Soc.} {\bf 167} 31P
\item[] Frey S and Gurvits L I 1977 {\it Vistas in Astronomy} {\bf 41} 271
\item[] Frey S, Gurvits L I, Kellermann K I, Schilizzi R T and Pauliny-Toth I I K 1997
        {\it Astron. Astrophys.} {\bf 325} 511
\item[] Frey S, Mosoni L, Paragi Z and Gurvits L I 2003 {\it Mon. Not. R. Astron. Soc.}
        {\bf 343} L20
\item[] Gilliland R L and Phillips M M 1998 {\it IAU Circ.} 6810
\item[] Gurvits L I 1994 {\it Astrophys. J.} {\bf 425} 442
\item[] Gurvits L I, Schilizzi R T, Barthel P D, Kardashev N S, Kellermann K I, Lobanov A P,
        Pauliny-Toth I I K and Popov M V 1994 {\it Astron. Astrophys.} {\bf 291} 737
\item[] Gurvits L I, Kellermann K I and Frey S 1999 {\it Astron. Astrophys.} {\bf 342}                    378
\item[] Hanany S \etal 2000 {\it Astrophys. J.} {\bf 545} L5
\item[] Hoyle F 1959 {\it Proc. IAU Symp. 9: Paris Symposium on Radio Astronomy}
        (Stanford: Stanford University Press) p 529
\item[] Jackson J C 1973 {\it Mon. Not. R. Astron. Soc.} {\bf 162} 11P
\item[] Jackson J C and Dodgson M 1996 {\it Mon. Not. R. Astron. Soc.} 278, 603
\item[] Jackson J C and Dodgson M 1997 {\it Mon. Not. R. Astron. Soc.} 285, 806
\item[] Jain D, Dev A and Alcaniz J S 2003 {\it Class. Quant. Grav.} {\bf 20} 4485
\item[] Kembhavi A K and Narlikar J V 1999 {\it Quasars and Active Galactic Nuclei}
        (Cambridge: Cambridge University Press) chapter 9 
\item[] Kellermann K I 1972 {\it Astronom. J.} {\bf 77} 531
\item[] Kellermann K I 1993 {\it  Nature} {\bf 361} 134
\item[] Kellermann K I 1994 {\it  Aust. J. Phys.} {\bf 47} 599
\item[] Kormendy J 2001a {ASP Conf. Ser. 230: Galaxy Disks and Disk Galaxies}
        (San Francisco: Astronomical Society of the Pacific) p 247
\item[] Kormendy J 2001b {\it Rev. Mex. Astronom. Astrofís. (Serie de Conferencias)}
        {\bf 10} 69
\item[] Lasenby A N, Bridle S L and Hobson M P 2000
        {\it Astrophys. Lett. \& Communications} {\bf 37} 327
\item[] Legg T H 1970 {\it  Nature} {\bf 226} 65
\item[] Lind K R and Blandford R D 1985 {\it Astrophys. J.} {\bf 295} 358
\item[] Livio M 2001 {\it Proc. STScI Symp. 13: Supernovae and Gamma-Ray Bursts}
        (Cambridge: Cambridge Univ. Press) p 334
\item[] Lima J A S and Alcaniz J S 2002 {\it Astrophys. J.} {\bf 566} 15
\item[] Masson C R 1980 {\it Astrophys. J.} {\bf 242} 8
\item[] Miley G K, 1971 {\it Mon. Not. R. Astron. Soc.} {\it 152} 477
\item[] Nilsson K, Valtonen M J, Kotilainen J and Jaakkola T 1993
        {\it Astrophys. J.} {\bf 413} 453
\item[] Orr M J L and Browne I W A 1982 {\it Mon. Not. R. Astron. Soc.} {\bf 200} 1067
\item[] Ostriker J P and Steinhardt P J 1995 {\it  Nature} {\bf 377} 600
\item[] Padovani P and Urry C M 1990 {\it Astrophys. J.} {\it 356} 75
\item[] Paragi Z, Frey S, Gurvits L I, Kellermann K I, Schilizzi R T, McMahon R G,
        Hook I M and Pauliny-Toth I I K 1999 {\it Astron. Astrophys.} {\bf 344} 51
\item[] Peebles P J E 1984 {\it Astrophys. J.} {\bf 284} 439
\item[] Perlmutter S \etal 1999 {\it Astrophys. J.} {\bf 517} 565
\item[] Press W H, Flannery B P, Teukolsky S A and Vetterling W T 1986,
        {\it Numerical Recipes} (Cambridge: Cambridge University Press) pp 532--536
\item[] Preston R A, Morabito D D, Williams J G, Faulkner J, Jauncey D L,
        Nicolson G 1985 {\it Astronom. J.} {\bf 90} 1599
\item[] Readhead A C S, Taylor G B, Xu W, Pearson T J, Wilkinson P N, Polatidis A G 1996a
        {\it Astrophys. J.} {\bf 460} 612
\item[] Readhead A C S, Taylor G B, Pearson T J, Wilkinson P N 1996b
        {\it Astrophys. J.} {\bf 460} 634
\item[] Rees M 1966 {\it Nature} {\bf 211} 468
\item[] Richter G M 1973 {\it Astrophys. Lett.} {\bf 13} 63
\item[] Riess A G, Filippenko A V, Challis P, Clocchiatti A, Diercks A, Garnavich P M,
        Gilliland R L, Hogan C J, Jha S, Kirshner R P, Leibundgut B, Phillips M M, Reiss D,               Schmidt B P, Schommer R A, Smith R C, Spyromilio J, Stubbs C, Suntzeff N B and
        Tonry J 1998 {\it Astronom.  J.} {\bf 116} 1009
\item[] Riess A G, Nugent P E, Gilliland R L, Schmidt B P, Tonry J, Dickinson M,
        Thompson R I, Budavári T, Casertano S, Evans A S, Filippenko A V, Livio M,
        Sanders D B, Shapley A E, Spinrad H, Steidel C C, Stern D, Surace J and
        Veilleux S 2001 {\it Astrophys. J.} {\bf 560} 49
\item[] Scheuer P G A and Readhead A C S 1979 {\it Nature} {\bf 277} 182
\item[] Schmidt B P \etal 1998 {\it Astrophys. J.} {\bf 507} 46
\item[] Shklovsky I S 1964a {\it Soviet Astron.--AJ} {\bf 7} 748
\item[] Shklovsky I S 1964b {\it Soviet Astron.--AJ} {\bf 7} 972
\item[] Singal A K 1988 {\it Mon. Not. R. Astron. Soc.} {\bf 233} 87
\item[] Spergel D N, Verde L, Peiris H V, Komatsu E, Nolta M R, Bennett C L, Halpern M,
        Hinshaw G, Jarosik N, Kogut A, Limon M, Meyer S S, Page L, Tucker G S,
        Weiland J L, Wollack E and Wright E L 2003
        {\it Astrophys. J. Suppl.} {\bf 148} 175
\item[] Taylor G B, Vermeulen R C, Readhead A C S, Pearson T J, Henstock D R,
        and Wilkinson P N 1996 {\it Astrophys. J. Suppl.} {\bf 107} 37 
\item[] Thompson A R, Moran J M and Swenson G W Jr 1986
        {\it Interferometry and Synthesis in Radio Astronomy} (New York: Wiley) p 13
\item[] Turner M S, Steigman G and Krauss L L 1984 {\it Phys. Rev. Lett.} {\bf 52}                        2090
\item[] Vishwakarma R G 2001 {\it Class. Quant. Grav.} {\bf 18} 1159
\item[] Wardle J F C and Miley G K 1974 {\it Astron. Astrophys.} {\bf 30} 305
\item[] Wiik K and Valtaoja E 2001 {\it Astron. Astrophys} {\bf 366} 1061
\item[] Wilkinson P N, Polatidis A G, Readhead A C S, Xu W and Pearson T J, 1994
        {\it Astrophys. J.} {\bf 432} L87
\item[] Zhu Z--H and Fujimoto M--K 2002 {\it Astrophys. J.} {\bf 581} 1

\endrefs

\newpage
\section{Figures}

\end{document}